\documentclass{wssci}
\usepackage{verbatim}
\usepackage{indentfirst}
\usepackage{epstopdf}
\usepackage{subfigure}
\usepackage{color}
\begin{document}
\title{ Towards using general purpose graphics processing unit (GPGPU) units for accelerating the batched perfectly stirred reactor (PSR) calculations }
\author{
%
Sudip Adhikari$^{1,^{\ast}}$, Alan Sayre$^2$, and Abhilash J. Chandy$^3$
}
\date{
%
$^1$Department of Mechanical Engineering,
 University of Akron,
Akron, OH - 44325-3903\\[6pt]
$^2$Babcock and Wilcox Power Generation Group,Inc., 
20 South Van Buren Avenue, Barberton, OH 44203-0351\\[6pt]
$^3$Department of Mechanical Engineering, Indian Institute of Technology Bombay, Mumbai, Maharashtra, India
\\
$^{\ast}$Corresponding Author Email: sa112@zips.uakron.edu
}
\maketitle
\begin{abstract}
Detailed analysis of efficiency and pollutant emission characteristics of practical turbulent combustion devices using complex combustion kinetics often depend on the interactions between the combustion chemistry involving both gasses species and soot, and turbulent flow characteristics. Modeling of such combustion system often requires the use of chemical kinetic mechanisms with hundreds of species and thousands of reactions. Perfectly stirred reactors (PSR) are idealized reactor environments, where the reacting species have high rate of stirring, and the combustion products are uniformly distributed inside the reactor. PSRs have been found very useful in the study of flame stabilization, prediction of pollutants such as $\textrm{NO}_{x}$ formation, development and testing chemical reaction mechanisms, and investigation of soot formation and growth.
The fundamental equations describing a PSR constitute systems of highly nonlinear algebraic equations, due to the complex relationship between the net production rate of the species and the species concentration, which ultimately makes the equations stiff, and the solution of such equations become highly compute-intensive leading to the need for a efficient and robust solution algorithms.
Graphics processing units (GPUs) have widely been used in the past as a cost-effective alternate to central processing units (CPUs), and highly parallel threads of GPUs can be used in a efficient manner to improve the performance of such algorithms for speeding up the calculations. A highly parallelized GPU implementation is presented for a batched calculation of PSR model, using a robust and efficient non-linear solver for gas phase chemical reactions and is further coupled to one of the widely used moment methods of solutions of soot equations, the Method of Moment with Interpolative Closure (MOMIC). 
\end{abstract}

\section{Introduction}

Accurate prediction of pollutant and gaseous byproducts emanating from practical combustion devices due to chemical reactions has been an increasingly important research topic in the recent years.  Unlike homogeneous thermodynamic systems, whose chemical and physical properties are uniform throughout the system, and whose simulations are usually in the order of a few seconds, practical combustion devices in the industrial combustion systems are inhomogeneous thermodynamic systems, and the simulations of such systems becomes complicated, because of the strong interaction between chemistry, turbulent flow, heat transfer and mass transfer.  Detailed simulation of such combustion systems remains computationally expensive and almost infeasible, despite the advancement in CPU hardware.  So there has also been an enormous effort for the development of efficient solution strategies by exploring the underlying theoretical and computational aspects related to detailed chemical reaction mechanisms.  Two of the approaches for reducing the central processing unit (CPU) time include the use of storage-retrieval techniques such as \textit{in-situ} adaptive tabulation (ISAT) \cite{pope1997}, and reduced kinetic mechanisms \cite{bhattacharjee2003optimally}, but, both of these approaches incur CPU time savings at the expense of accuracy.  There is a need for a strategy that can save CPU time without losing the accuracy of the predicted solutions. The computational parallelization of the calculations can be a reasonable alternative, yet the parallelization algorithm and the choice of parallel computing hardware architecture has a significant impact on the effectiveness of the algorithm and speedup achieved.

Perfectly stirred reactor is considered as an ideal reactor with high rate of stirring \cite{GKGM} inside reactor control volume, leading to the perfect mixing of reactants. Because of the intense recirculation inside the reactor, reactants and products rapidly mix to form a homogeneous mixture. PSRs have been used in the past for the study of soot formation and growth \cite{adhikari2015,adhikari2016hybrid,adhikari2013aps,adhikari2017situ,adhikari2018situ}, and chemical kinetics \cite{BDC1973}. In the case of a PSR, the Damkohler number, \textit{Da}, which is the ratio of the rate of chemical reaction to the rate of mixing, is $\ll 1$, and the chemical reactions occur over a timescale that is much larger than the mixing timescale.
The properties inside the reactor are assumed to be spatially uniform, because of the efficient mixing of the species, thereby providing a homogeneous composition of gas inside the reactor. In another words, a PSR can be described by a single reactor temperature and concentration of the species. In the current study, the walls of the PSR are considered to be non-catalytic. There are no deposits on the reactor walls, and therefore the inlet mass flow and outlet mass flow rates are equal. The efficiency of the reactor is dependent on its volume and the volume flow rate or residence time, and it can be related to the chemical processes occurring inside. 

Like the concept of PSR, the eddy dissipation combustion model \cite{MH76}, is based on the infinitely fast
chemistry hypothesis and assumes that the reaction rate is controlled by
turbulent mixing.  A generalized formulation of the eddy dissipation model was proposed in order to take into account finite rate chemistry effects.  However, this model was not suited to treat detailed reaction mechanisms and failed with number of reactions exceeding three or four
\cite{galletti2006numerical}.  On the other hand, the eddy dissipation concept (EDC), which is an extension of the eddy dissipation model, allows detailed kinetics to be included in the calculations \cite{Magnussen81}. The model assumes that chemical reactions occur within the smallest turbulent structures, called fine structures. In other words, these are treated as PSRs which exchange chemical species with the surrounding. The overall reaction rate in each PSR is controlled by chemical kinetics.  Conservation of mass, energy, and species govern a well stirred reactor yielding systems of nonlinear algebraic equations, the number of which, is equal to the number of species undergoing reactions inside. The larger the dimension of combustion system, the larger the number of PSRs, and hence the size of the system of nonlinear algebraic equations to be solved will also be larger.  The methods to solve such equations can be classified either as direct methods or iterative methods.  Direct methods involving LU decomposition and back solving are the most widely used methods, because of its simplicity, stability and robustness \cite{botsch2010polygon}.
Governing equations of the perfectly stirred reactor constitute a set of nonlinear-algebraic equations, which are solved by employing the Newton's iteration, and this in turn needs the calculation of Jacobian matrix for the purpose of linearization of nonlinear system.
The Jacobian matrix needed for the LU decomposition is obtained by applying finite difference approximation, which requires the calculation of net production rate of species evaluated from the reaction rate constants. 

While using EDC model, if the system being simulated has $n$ computational cells, and the reaction mechanism being used has $m$ reactions, a single iteration for such calculation consists of $n \times m$ evaluations of rate constant, and since the rate constant is only dependent on the temperature, the calculations for $n$ cells can be performed as a set of $n$-batched PSR calculations with each of the batch evaluating $m$ chemical rate constant evaluations. Since the rate constant calculations involve multiplication and exponentiation, the tasks could be carried out separately and the efficiency can be improved further with parallelizations on Central Processing Units (CPUs) or Graphics Processing Units (GPUs). GPUs are an alternative approach of accelerator technology, that can potentially improve the performance of various algorithms. GPU performance already exceeds that of modern CPUs by one or two orders of magnitude on selected applications \cite{BCIMQ08} well suited to the GPU architecture. In addition to raw computing performance, GPUs offer similarly impressive metrics for performance per dollar and performance per watt.  Also, massively-parallel, multi-core architecture of GPUs delivers higher performance, and allows researchers to tackle some of today's most demanding applications - applications that previously could only be handled by massive supercomputers or clusters, thus relieving to a great extent the burden on the environment or the power and cooling budget, mainly due to a large reduction in electrical and thermal requirements.  
GPUs have been used for a variety of studies in different areas in the past. Some of the them include linear algebra \cite{barrachina2008solving}, computational biology \cite{schatz2007high}, weather prediction \cite{michalakes2008gpu}, molecular dynamics \cite{le2013spfp}, and turbulence modeling \cite{rucki2014towards}. GPU related works in the field of combustion include the development of GPU enhanced algorithm for integration of stiff ordinary differential equation\cite{stone2013techniques,niemeyer2014accelerating,sewerin2015methodology}.
Stone\cite{stone2013techniques} implemented fourth order Runge-Kutta-Fehlberg ordinary differential equation solver in GPU and achieved a maximum speed up of 20.2x over CPU. The work of Niemeyer and Sung\cite{niemeyer2014accelerating} include the computation of chemical kinetics of a hydrogen oxidation mechanism using the explicit fifth-order Runge-Kutta-Cash-Karp method. They have reported GPU speedups of 126x and 25x over the single- and six-core CPU version, respectively. Sewerin and Rigopoulos\cite{sewerin2015methodology} describe the implementation of three-stage/fifth-order implicit Runge-Kutta method Radau5\cite{wanner1991solving} using OpenCL. They tested the implementation in the context of a transient equilibrium scheme for the flamelet model.
  
The main objective of this paper is to develop a general purpose GPU (GPGPU) implementation of rate constant calculation for a single iteration in the context of batched calculations of PSRs using CUDA, to investigate the soot distribution inside a PSR using a statistical state-of-the-art soot model such as the Method of Moment with Interpolative Closure (MOMIC) \cite{FR}.  

\section{Governing Equations}
The PSR is governed by three different conservation equations: conservation of mass, conservation of species and conservation of energy. The rate of change of total mass inside the PSR can be related to the mass inflow rate and mass outflow rate as:
\begin{equation}
\frac{d(\rho V)}{dt} = \dot{m}_{in}-\dot{m}_{out} 
\label{eqn:mass}
\end{equation}
The rate of change of concentration of mass fraction of a species inside the
PSR can be related to its molar production rate and
initial mass fraction as:
\begin{equation}
 \frac{dY_{i}}{dt} = \frac{\dot{m}_{in}}{\rho V} (Y_{i,in}-Y_{i}) +  \frac{\dot{\omega}_{i}^g MW_{i}}{\rho }
 \label{eqn:species}
\end{equation}
Finally, the energy equation is given by
\begin{equation}
 C_{p,mix}\frac{dT}{dt} = \frac{\dot{m}}{\rho V} \sum\limits_{i=1}^n Y_{i,in}(h_{i,in}-h_{i}) -
 \frac{\sum\limits_{i=1}^n h_{i}\dot{\omega}_{i}^g MW_{i}}{\rho } + \frac{\dot{Q}}{\rho V}-
 \frac{1dP}{\rho dt}
 \label{eqn:energy}
\end{equation}
\section{Computational formulation of Rate evaluation }

The species conservation equations governing a PSR are a set of nonlinear algebraic equations. Since these equations are nonlinear in nature, they have to be solved iteratively.  For this purpose the Newton Iteration technique is employed . In this method, one of the first steps involves constructing an error vector by substituting the initial guesses into the nonlinear-algebraic equations. The next step is to construct a Jacobian matrix. The Jacobian matrix is decomposed into
lower and upper triangular matrices, and then a back-solving operation is carried out to obtain the final solution. The PSR equations are solved by employing a hybrid/newton time integration approach as described and developed for a PSR with soot moment equations in one of the authors' previous studies \cite{adhikari2016hybrid}.  The hybrid approach  consists of two steps: a steady state Newton iteration and a pseudo-transient time integration. If convergence is not achieved using the Newton iteration, the solution estimate is conditioned using the time integration. The process of switching between the Newton iteration and time integration approaches continues until convergence is achieved, which is based on very specific criteria related to absolute and relative tolerances.  See \cite{adhikari2016hybrid} for more details.

The net production rate $\dot{\omega}_{i}^g$ in the equation \ref{eqn:species} needs to be calculated at every time instant, the Jacobian matrix is calculated, and is represented as the summation of the rate of progress variable for all reactions involving the $i^{th}$ species:
\begin{equation}
\dot{\omega}_{i}^g = \sum_{k=1}^{I} \nu_{i,k} q_{k} \qquad \quad (i=1,2,3,...n)
\label{eqn:omega_dot_def}
\end{equation}
where
\begin{equation}
\nu_{i,k} = \nu^{''}_{i} - \nu^{'}_{i}
\nonumber 
\label{eqn:nu_ki}
\end{equation}
and, $I$ is the total number of elementary reactions. $\nu^{''}_{i}$ and $\nu^{'}_{i}$ are the stoichiometric coefficient of the species, $i$, in the product and reactant side, respectively. The rate of progress variable $q_{k}$ for the reaction, $k$, is given by the difference of the forward rates and reverse rates as,
\begin{equation}
q_{k} = k_{f,k} \prod_{i=1}^{k} \big[X_{i}\big]^{\nu^{'}_{i,k}} -  k_{r,k} \prod_{i=1}^{k} \big[X_{i}\big]^{\nu^{''}_{i,k}}
\label{eqn:qk}
\end{equation}
where $X_{i}$ is the molar concentration of the species, $i$ and $k_{fk}$ and $k_{rk}$ are the forward and reverse rate constants of the $k^{th}$ reaction. The forward rate constant for the $I$ reactions are generally assumed to have the following Arrhenius temperature dependence:
\begin{equation}
k_{f,k} =  A_{k}T^{\beta_{k}}\text{exp}\Big( \frac{-E_{k}}{RT}\Big)
\label{eqn:forward_rate_expression}
\end{equation}
where $A_{k}$ is pre-exponential factor, $\beta_{k}$ is temperature exponent , and $E_{k}$ is the activation energy for reaction $k$.
 The reverse rate constants, $k_{r,k}$, are related to the forward rate constants through the equilibrium constants as follows:
\begin{equation}
 k_{r,k} = \frac{k_{f,k} }{K_{c,k} }
 \label{eqn:reverse_rate_expression}
\end{equation}
 The constant, $K_{c,k}$, for reaction, $k$, given by;
 \begin{equation}
K_{c,k} =  K_{p,k}\Big(\frac{P_{atm}}{RT}\Big)^{\sum_{i=1}^{n} \nu_{i,k}}
\label{eqn:Kck_expression}
\end{equation}
The equilibrium constants $K_{p,k}$ are obtained with the relationship,
\begin{equation}
K_{p,k} =  exp\Big(\sum_{i=1}^{n} \nu_{i,k}\frac{S^{o}_{i}}{R} - \sum_{i=1}^{n} \nu_{ik}\frac{H^{o}_{i}}{RT}\Big)
\label{eqn:equilibrium_constant_expression}
\end{equation}

For chemical mechanisms with a large number of reactions, calculation of the forward rate constants (in Equation \ref{eqn:forward_rate_expression}) is time consuming, specifically due to the exponential term.
Since the values of the pre-exponent factor, temperature exponent and the activation energy are different for the different reactions, the forward rate expression for different reactions can be evaluated separately and independently, and hence are better suited for GPU parallelization.
%
From Equations \ref{eqn:reverse_rate_expression}, \ref{eqn:Kck_expression} and \ref{eqn:equilibrium_constant_expression}, it can be seen that the reverse rate constants are dependent on the entropy, enthalpy and the stoichiometric coefficients of the species, both on the reactant and product sides, which clearly indicate that their calculations are not straightforward and involve a lot of operations. The first step in the calculation of reverse rate constants involves determining the equilibrium constant from Equation \ref{eqn:equilibrium_constant_expression}. The right hand side of the Equation \ref{eqn:equilibrium_constant_expression} is an exponential term, where the exponent involves two different summations. The summation runs over the total number of species involved in a single  reaction.  The more the number of reactions, the more time consuming this particular operation becomes, and thereby is more suitable for GPU parallelization. The second step is the calculation of the constant, $K_{c,k}$, which involves the multiplication of equilibrium constants with $\frac{P_{atm}}{RT}$ raised to the power of another summation term.  Again, the summation is over the number of species in a reaction and so is computationally expensive for a large chemical mechanisms, which in turn makes them suitable candidates for GPU parallelization as well.  The third and final step is the evaluation of expression from Equation \ref{eqn:reverse_rate_expression}, which is a division operation, and can easily be set up for GPU acceleration.

\section{GPU Implementation}

The pseudo code for the current GPU implementation is presented in this section. The \textit{CalcRateofProgress()} function is the main function which calls three Cuda kernel functions for the calculation of reaction rate constants.
Dynamic memory allocation and variable initialization is carried out in the beginning. Data required for the calculations is then transferred from the CPU to the GPU.  After the calculation of rate constants, GPU data is transferred back to CPU. During the course of these calculations, the stoichiometric coefficients of the gaseous reactant and product species have a constant value, and hence, they are stored as a constant memory on the GPU. 

function \textit{CalcRateofProgress()}\\
\-\hspace{0.25cm}\{\\
\-\hspace{0.5cm}call \textit{CalcFwdRateConst($k_{fk}, I, n, .....$)}\\
\-\hspace{0.5cm}call \textit{CalcKeqConst($k_{pk}, I, n, S^{o}_{i}, H^{o}_{i} .....$)}\\
\-\hspace{0.5cm}call \textit{CalcRevRateConst($k_{rk}, k_{fk}, K_{pk}, I, n, .....$)}\\
\-\hspace{0.5cm} transfer $k_{fk}$ and $k_{rk}$ from GPU to CPU\\
\-\hspace{0.5cm}calculate the rate of progress variable $q_{k}$\\
\-\hspace{0.25cm}\}\\
end function \textit{CalcRateofProgress()}\\

cuda kernel \textit{CalcFwdRateConst($k_{fk}, I, n, .....$)}\\
\-\hspace{0.25cm} for each GPU thread  in parallel\\
\-\hspace{0.5cm} calculate forward rate constant $k_{fk}$\\      
end cuda kernel \textit{CalcFwdRateConst()}\\

cuda kernel \textit{CalcKeqConst($k_{pk}, I, n, S^{o}_{i}, H^{o}_{i} .....$)}\\
\-\hspace{0.25cm} for each GPU thread in parallel\\
\-\hspace{0.5cm} calculate equilibrium constant $K_{pk}$ in parallel \\      
end cuda kernel \textit{CalcKeqConst()}\\

cuda kernel \textit{CalcRevRateConst($k_{rk}, k_{fk}, K_{pk}, I, n, .....$)}\\
\-\hspace{0.25cm} for each GPU thread in parallel\\
\-\hspace{0.5cm} calculate constant $k_{ck}$ in parallel \\
\-\hspace{0.5cm} calculate equilibrium constant $k_{rk}$ in parallel \\      
        end cuda kernel \textit{CalcRevRateConst()}\\


\section{Validation of Physical Results}

\begin{figure}[!h]
  \centering
  \subfigure{
  \includegraphics[width=0.45\textwidth]{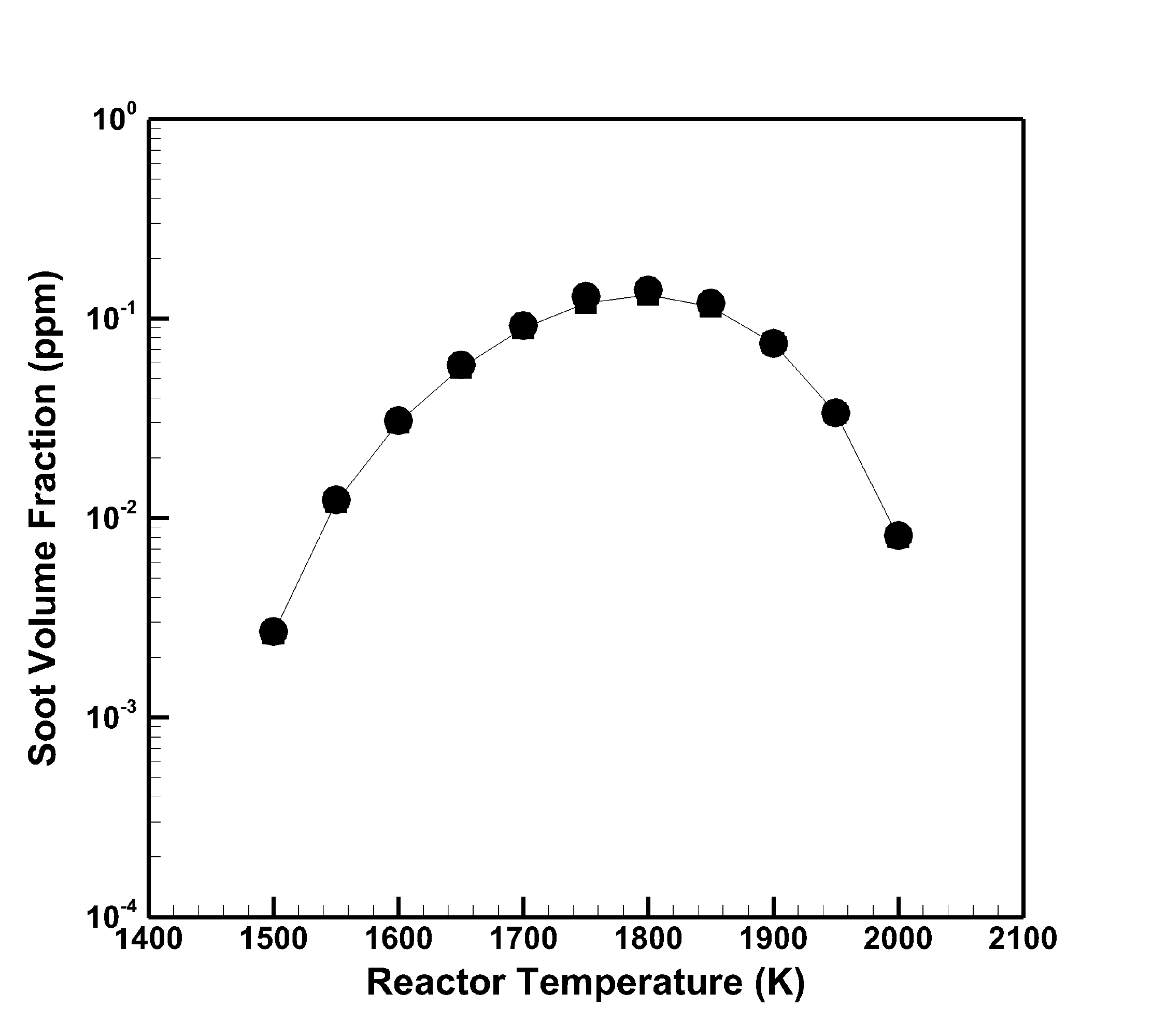}}
  \subfigure{
  \includegraphics[width=0.45\textwidth]{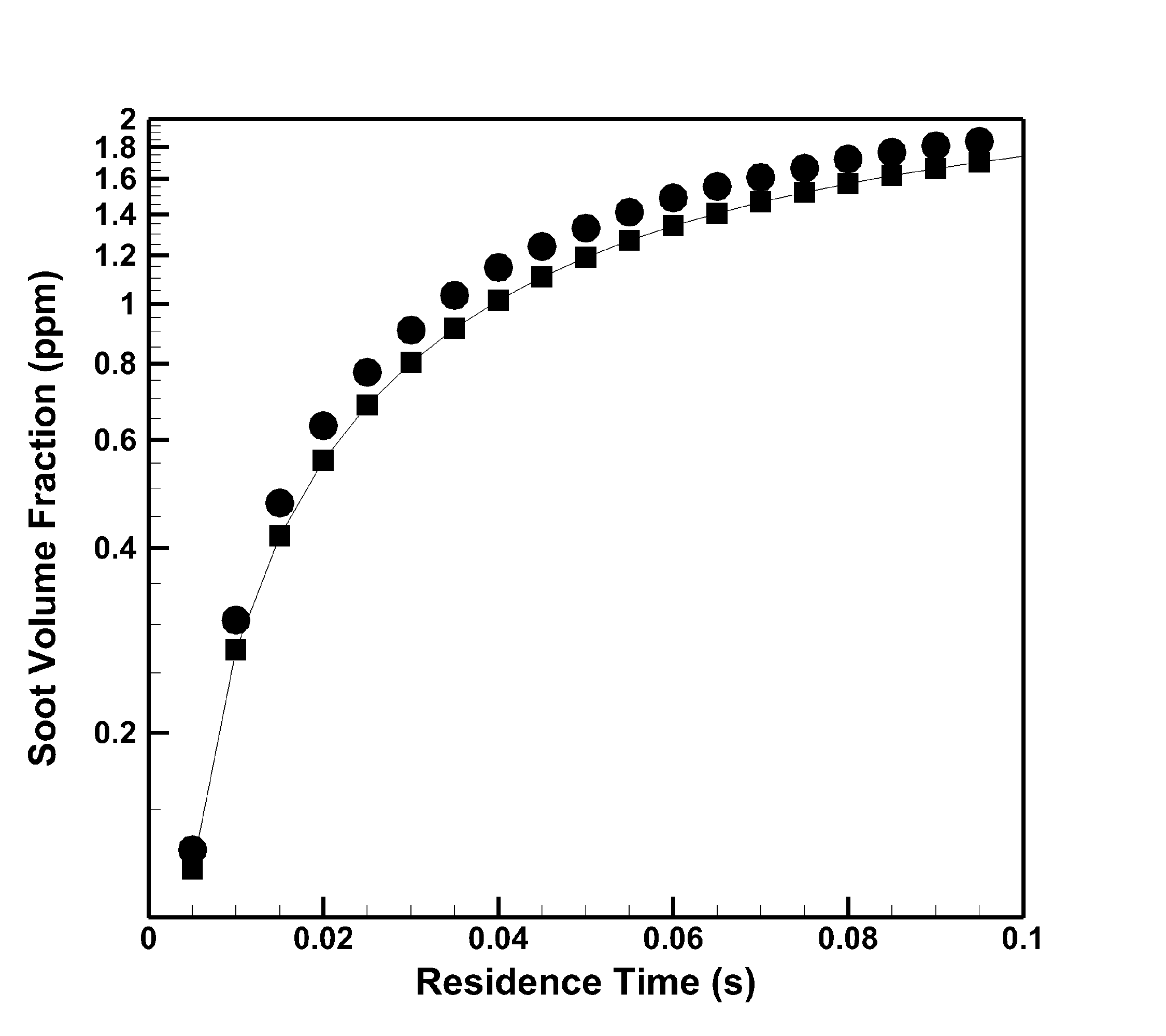}}  
  \caption{Variation of soot volume fraction with reactor temperature (left) and residence time (right) for ethylene/air combustion from PSR simulations; Results from Brown et al. - symbols, CPU Result with MOMIC - lines with closed square symbols, GPU Result with MOMIC - lines with open square symbols }
  \label{fig:ethylene_air_combustion_for_psr_validation}
\end{figure}

\begin{figure}[!h]
  \centering
  \subfigure{
  \includegraphics[width=0.45\textwidth]{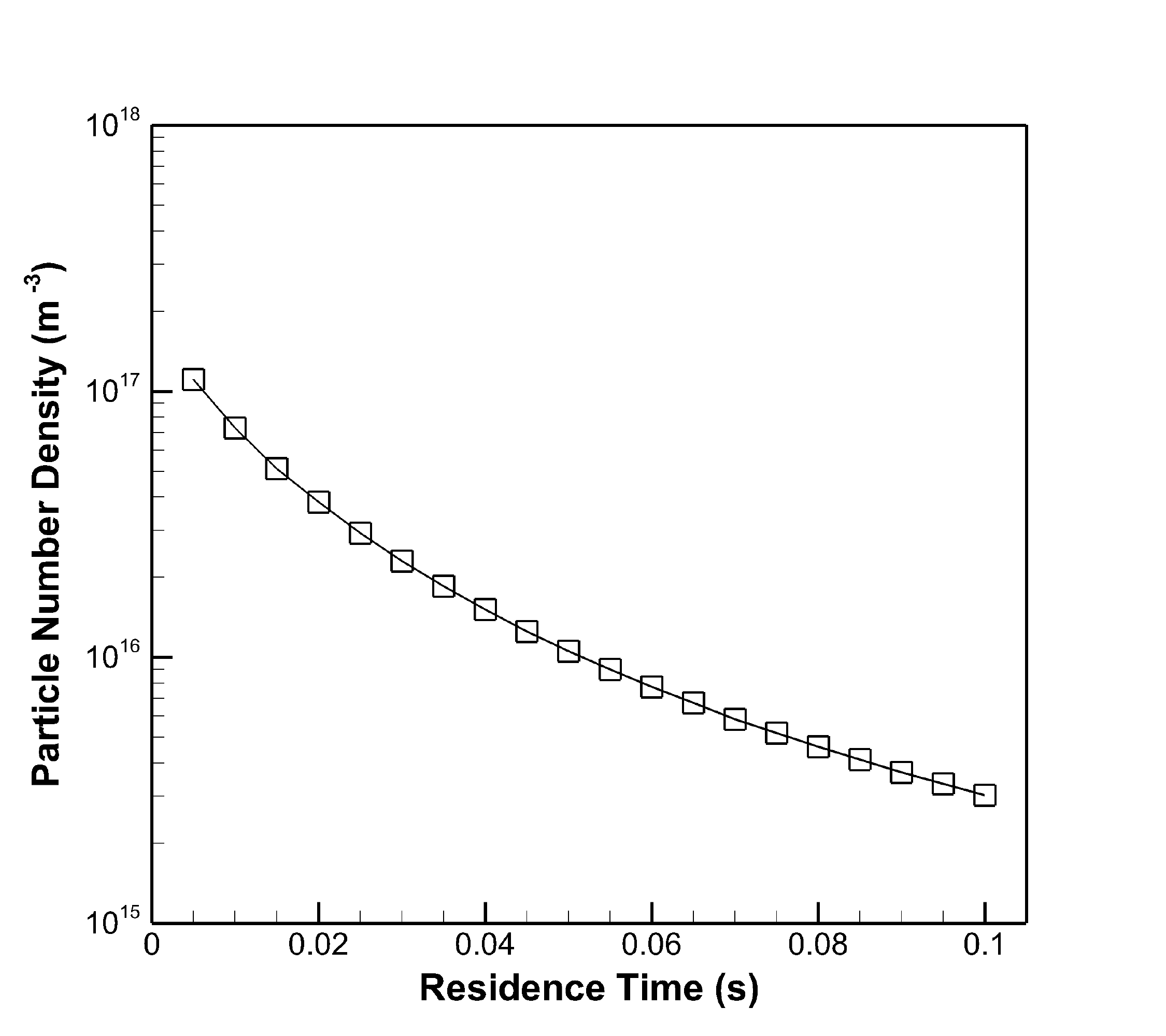}}
  \subfigure{
  \includegraphics[width=0.45\textwidth]{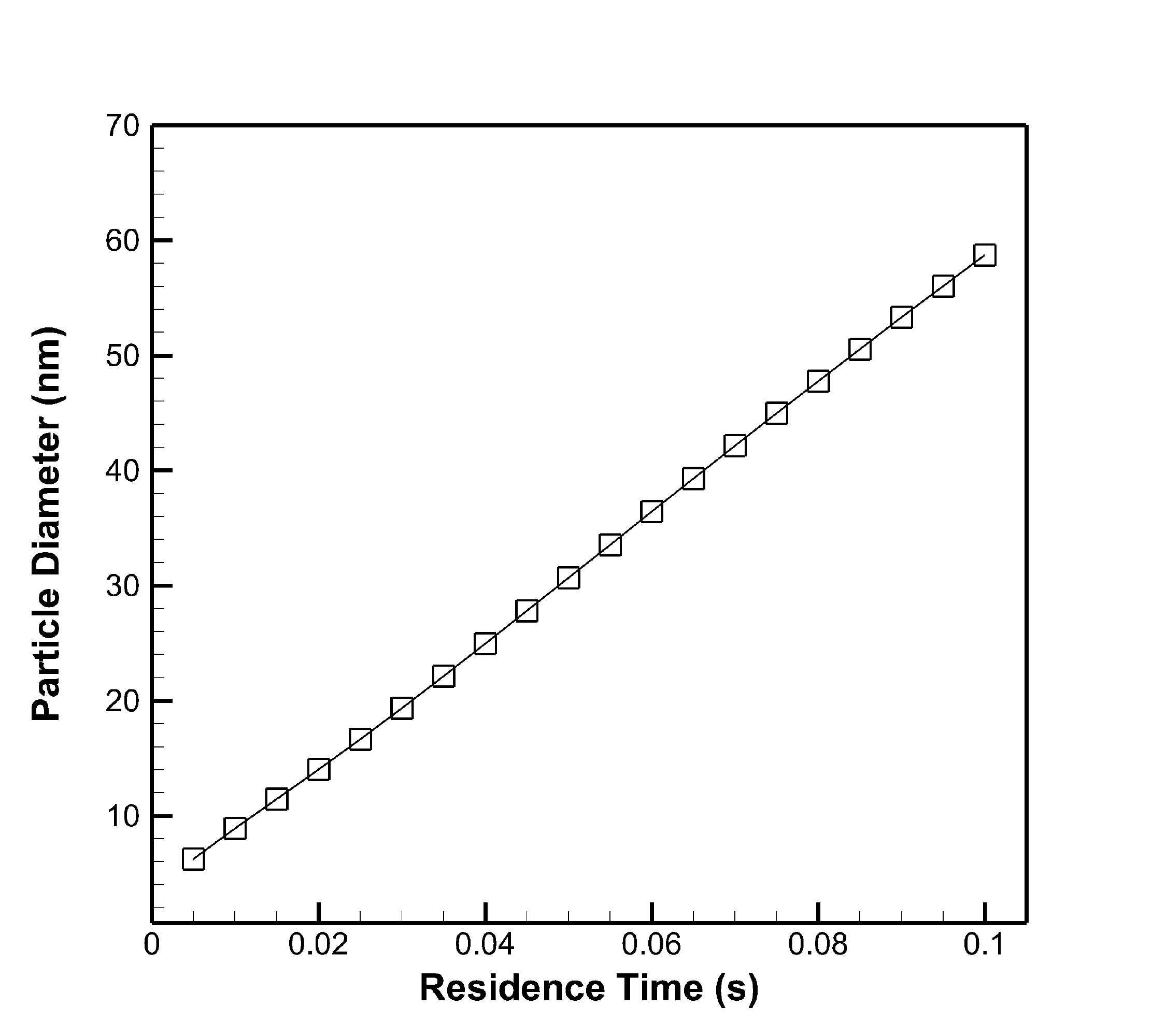}}
  \subfigure{
  \includegraphics[width=0.45\textwidth]{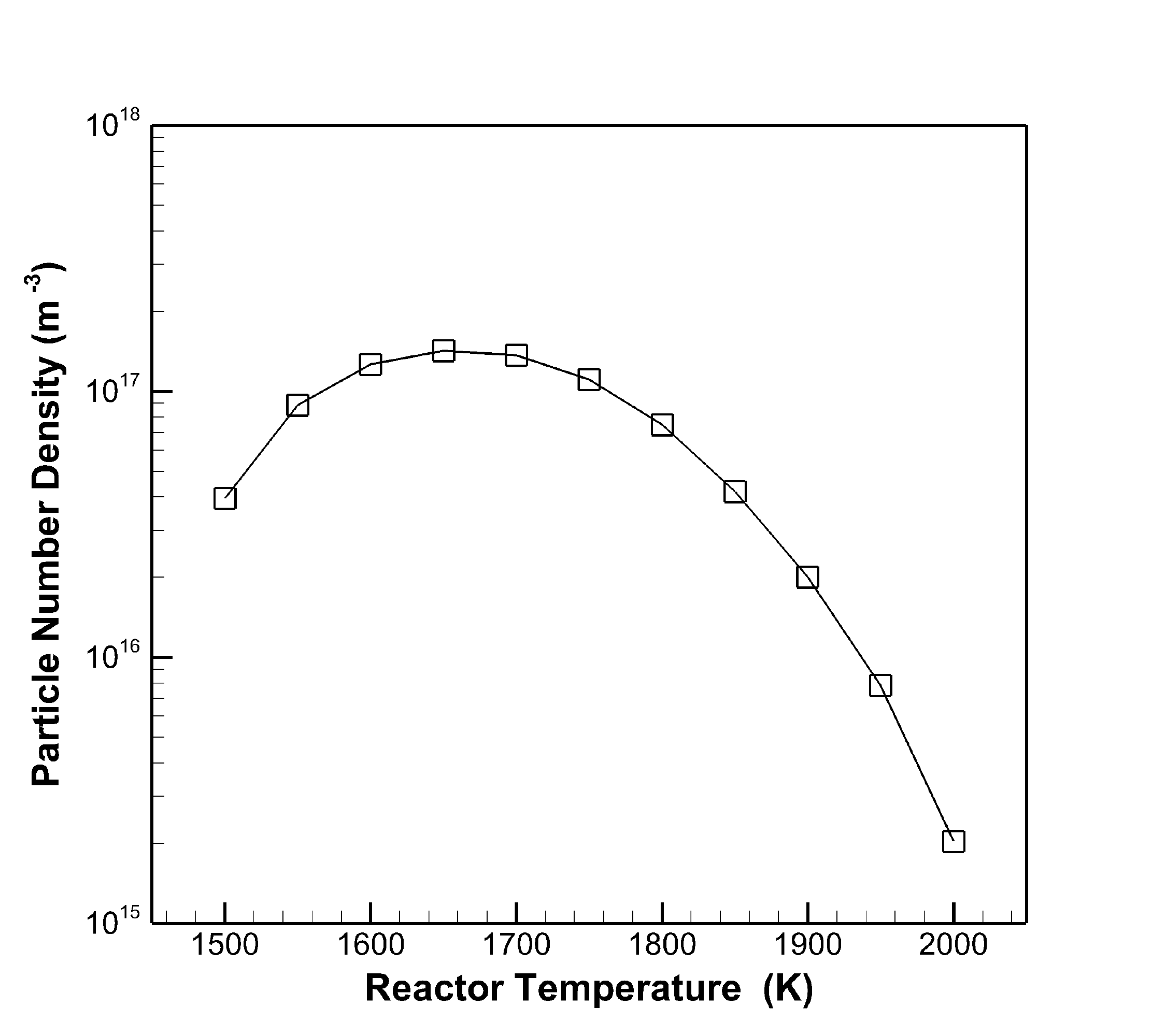}}
  \subfigure{
  \includegraphics[width=0.45\textwidth]{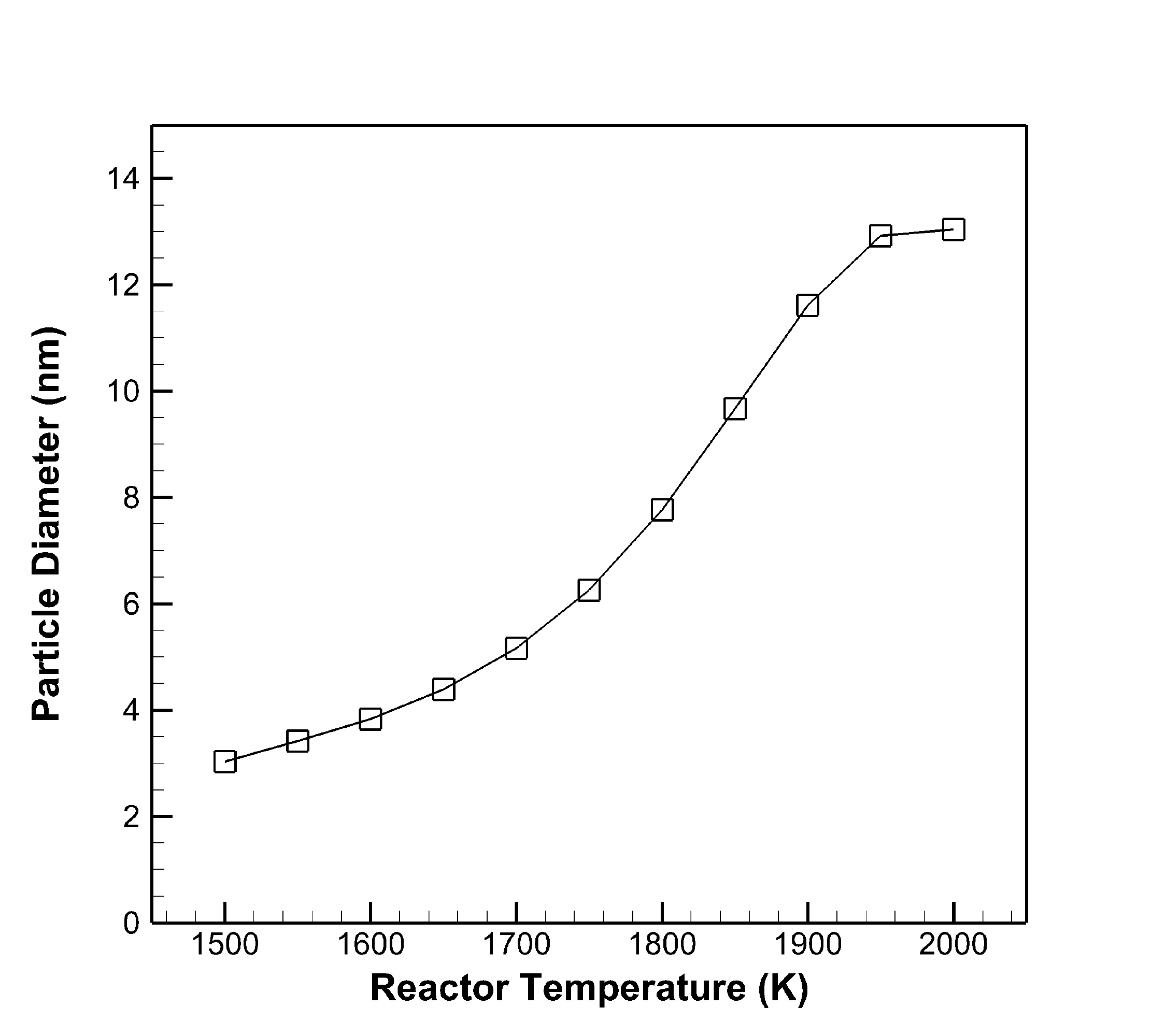}}  
  \caption{Variation of particle number density (top left) and particle diameter (top right) with residence time, and variation of particle number density (bottom left) and particle diameter (bottom right) with reactor temperature for ethylene/air combustion from PSR simulations; GPU Result with MOMIC - lines with open square symbols }
  \label{fig:PND_and_PD_vs_TAU_and_TEMP}
\end{figure}
This section presents the results for ethylene/air combustion in a PSR for the purpose of validating the in-house CPU code and the results from GPU implementation. Ethylene/air combustion in a PSR at specific operating conditions was previously numerically studied and presented by Brown et al.\cite{NKM1998}, which serves as a reference here in the current study.  Two sets of PSR simulations are carried out here: the first, with the reactor temperature varying from 1500 K to 2000 K in steps of 50 K, at a pressure of 1 atm, with a residence time of 0.005 s, and the second, with the reactor residence time varying from 5 ms to 95 ms in steps of 5 ms, again at a pressure of 1 atm, with a reactor temperature of 1750 K. The fuel-air (F/A) equivalence ratio ($\phi$) was 2.5 for all of the calculations presented in this section. Figure \ref{fig:ethylene_air_combustion_for_psr_validation} shows the variation of soot volume fraction with reactor temperature and residence time using MOMIC as soot model. Results from Brown et. al \cite{NKM1998} are also digitized and plotted alongside. The reaction mechanism used for the simulations was that of Wang and Frenklach \cite{wang1997detailed}.  Firstly from the figure, it is clear that, the soot volume fraction calculations from CPU code match perfectly with those calculated from GPU code, indicating the correctness of the GPU implementation.  Also, the soot volume fraction calculations at different residence times using the CPU code matches well with the results from Brown et. al \cite{NKM1998}, with a maximum difference being 0.1 ppm from \cite{NKM1998}. With regard to the reactor soot volume fraction versus reactor temperature, calculations from CPU code match perfectly with the predictions from Brown et. al \cite{NKM1998} and GPU code. Overall, the predictions from CPU code and GPU code compare very well with those from Brown et. al \cite{NKM1998}, thereby validating the CPU code and GPU implementation.

Figure \ref{fig:PND_and_PD_vs_TAU_and_TEMP} shows the variation of soot particle number density and soot particle diameter with reactor residence time, and also with reactor temperature. Soot particle number density decreases with residence time, due to the high coagulation rates at high residence times. The particle diameter on the other hand, increases almost linearly with residence time. With regard to reactor temperature, particle number density first increases, attains a maximum, and then decreases.  In this case, the soot particle diameter increases slowly for lower temperatures ($< 1750 K$) and shows a much steeper increases until 1950 K, after which there is no change.

\begin{figure}[!h]
  \centering
  \subfigure{
  \includegraphics[width=0.45\textwidth]{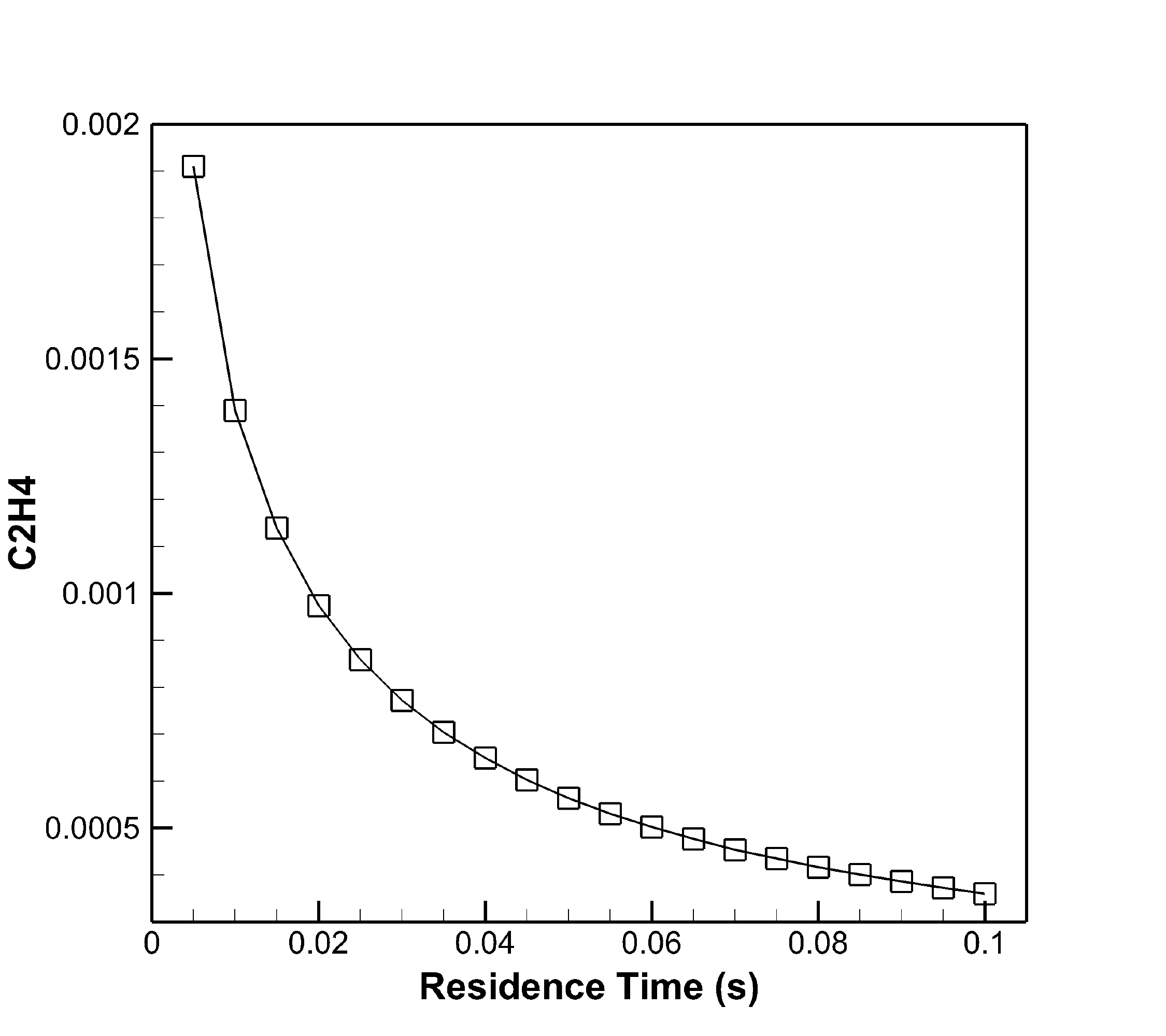}}
  \subfigure{
  \includegraphics[width=0.45\textwidth]{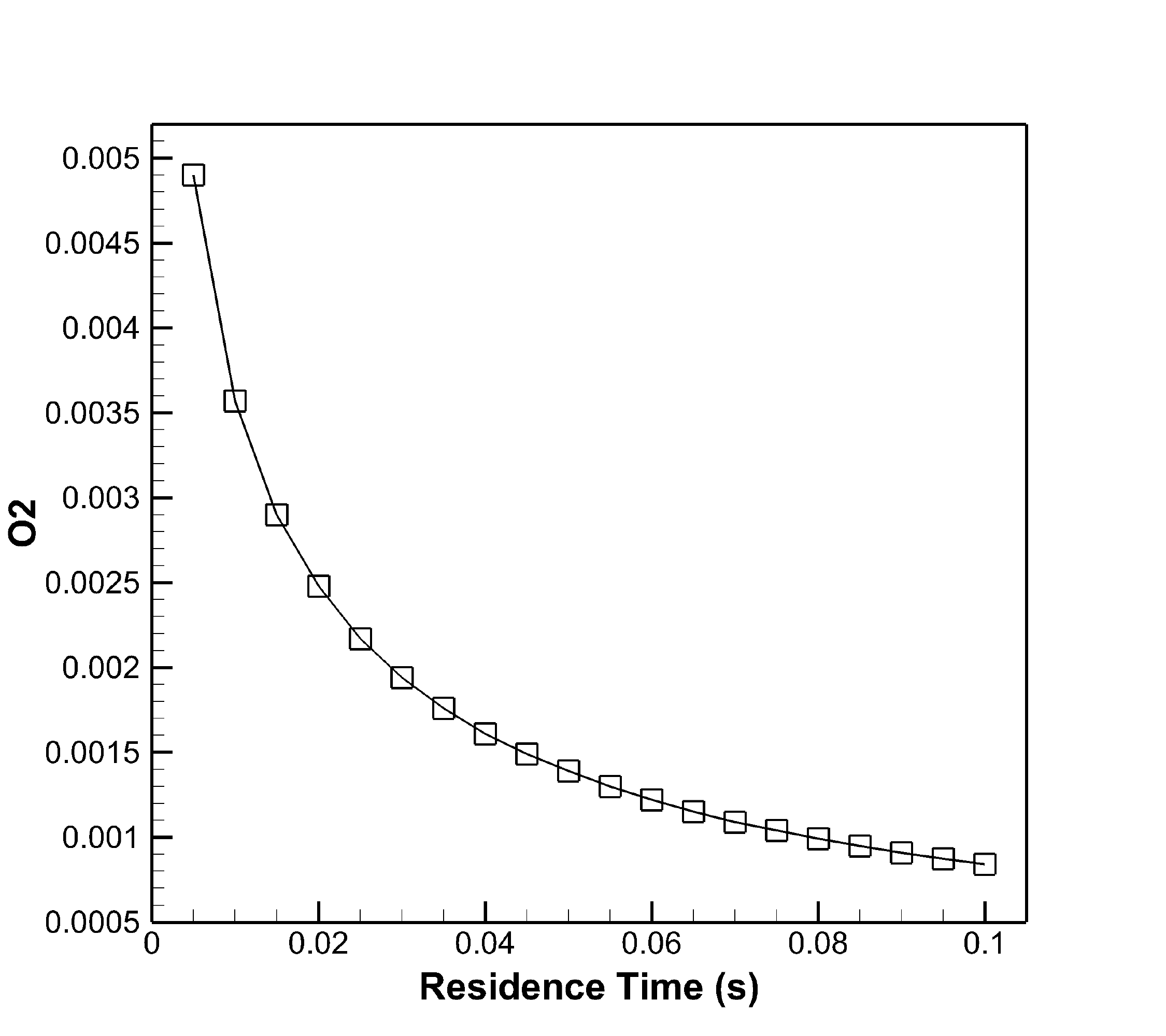}}
  \subfigure{
  \includegraphics[width=0.45\textwidth]{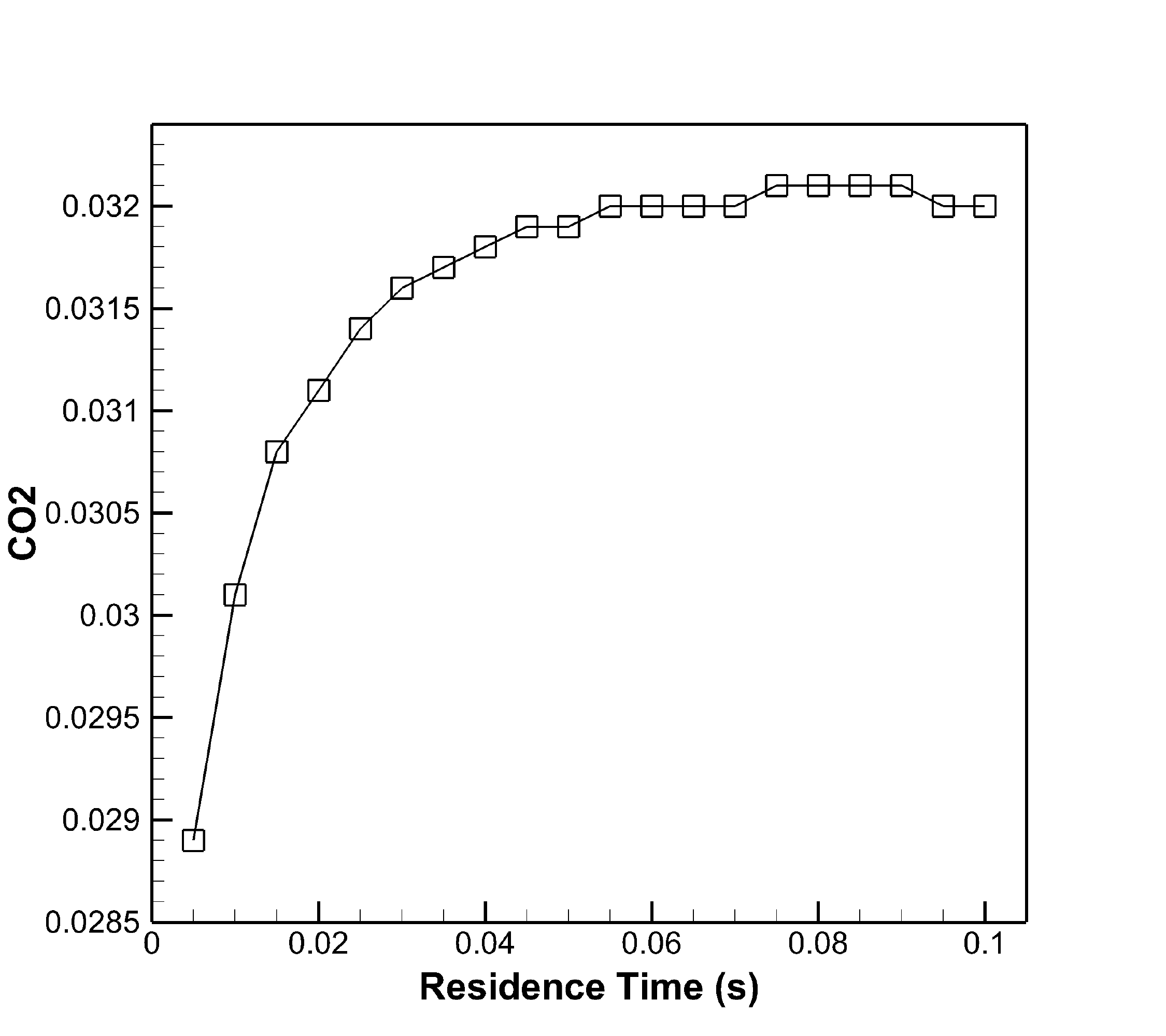}}
  \subfigure{
  \includegraphics[width=0.45\textwidth]{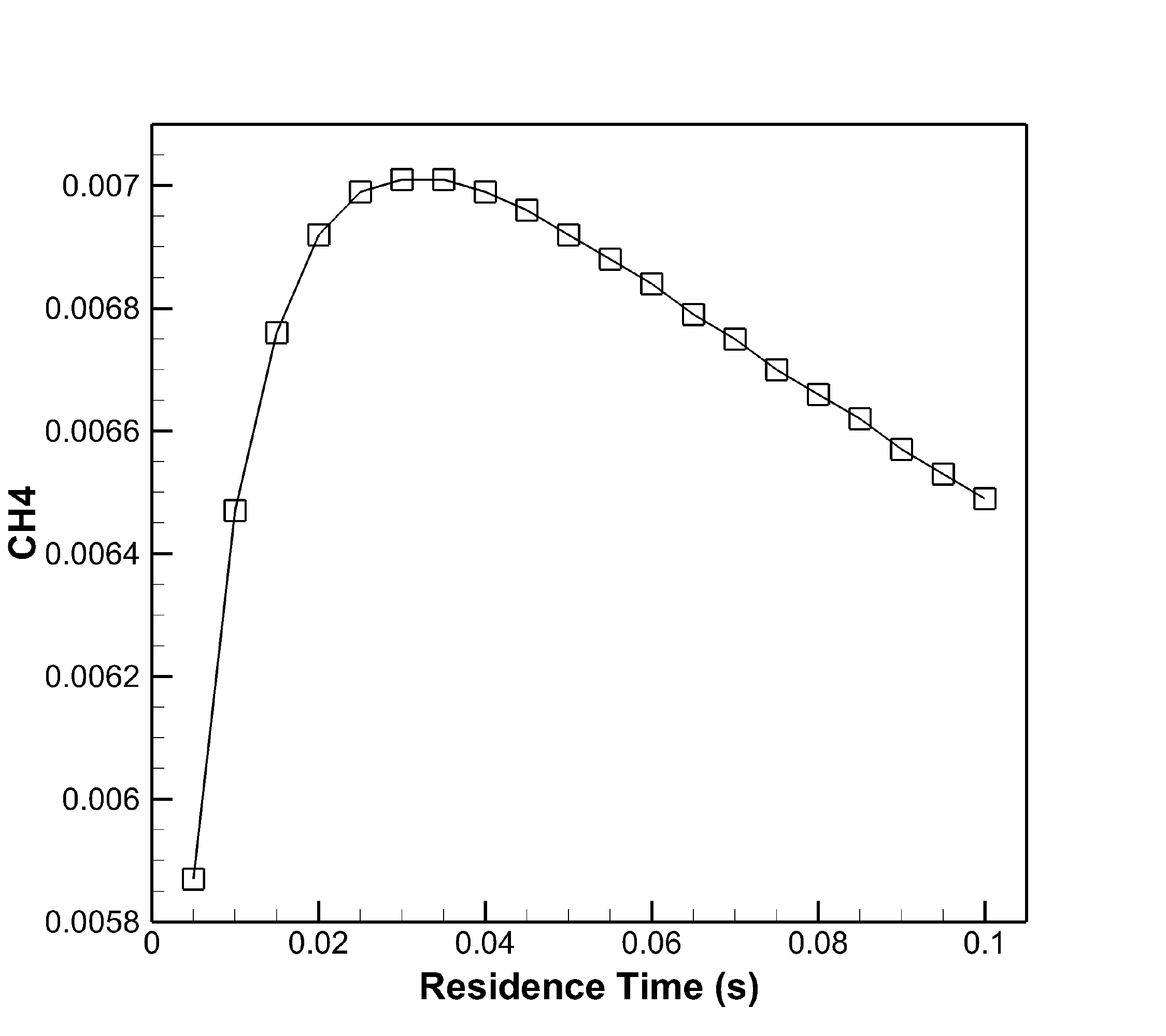}}  
  \caption{Mole fraction of C2H4 (top left), Mole fraction of O2 (top right), Mole fraction of CO2 (bottom left), Mole fraction of CH4 (bottom right) with  residence time for ethylene/air combustion from PSR simulations; GPU Result with MOMIC - lines with open square symbols }
  \label{fig:Species_mole_fraction_vs_TAU}
\end{figure}

\begin{figure}[!h]
  \centering
  \subfigure{
  \includegraphics[width=0.45\textwidth]{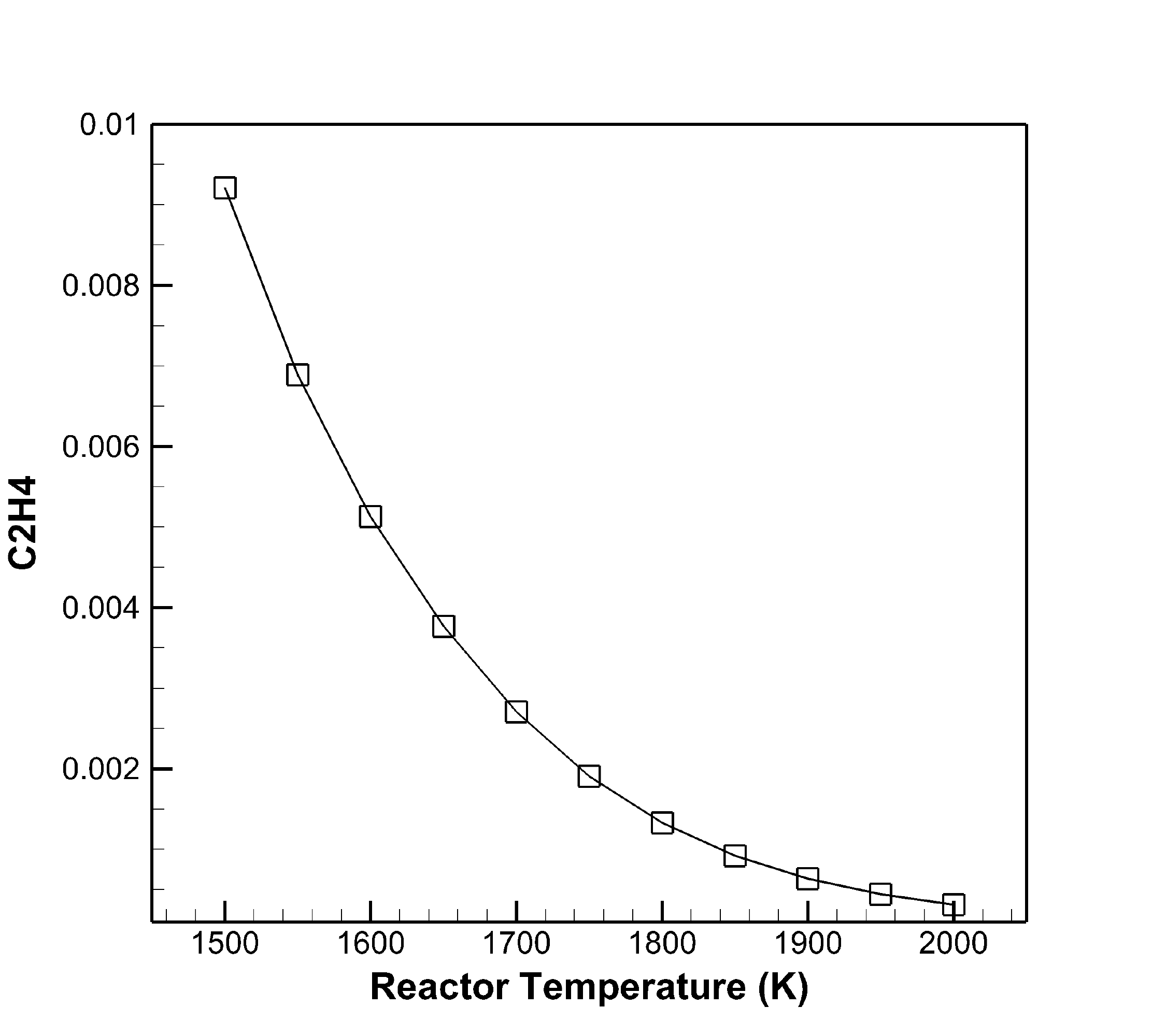}}
  \subfigure{
  \includegraphics[width=0.45\textwidth]{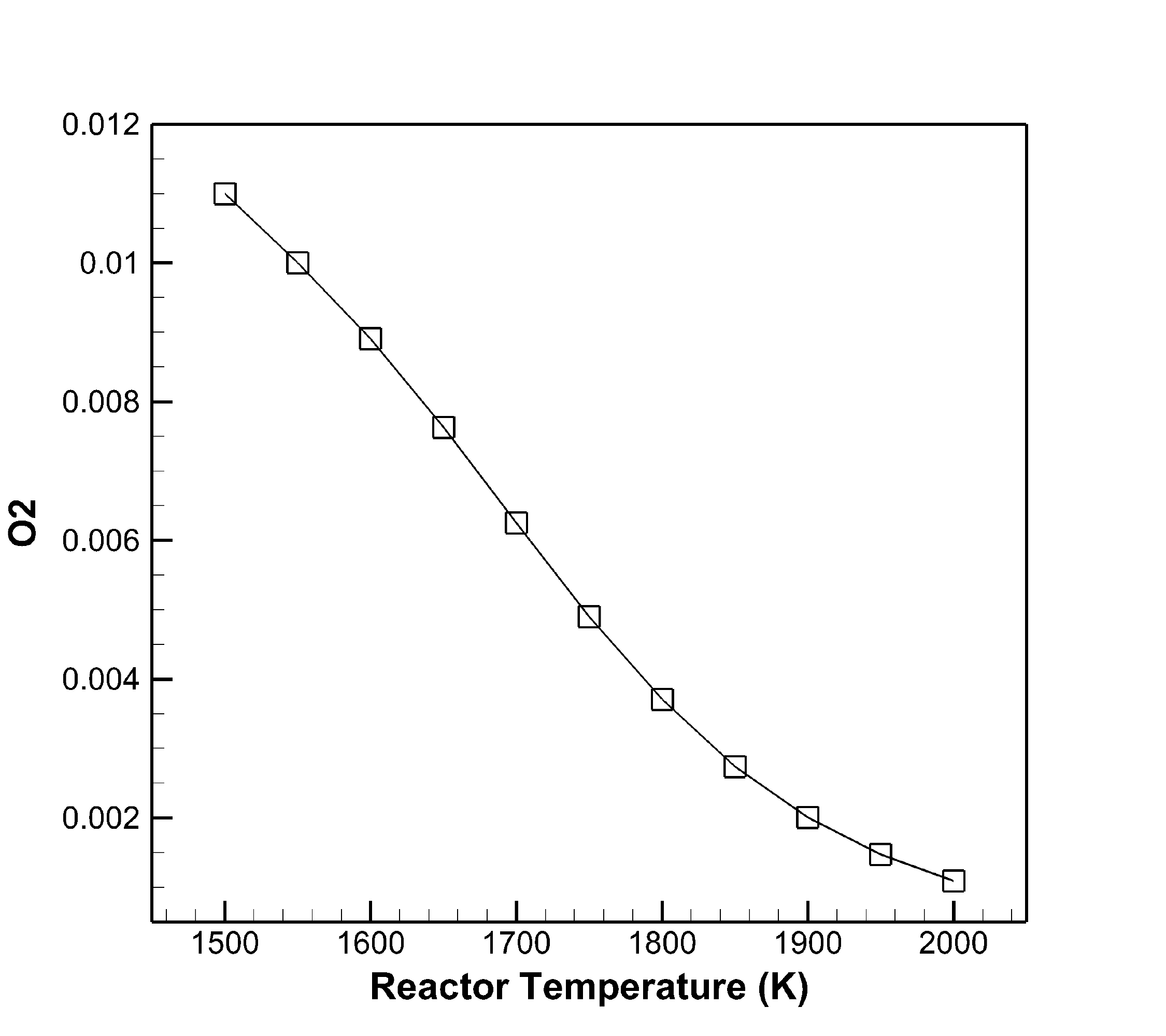}}
  \subfigure{
  \includegraphics[width=0.45\textwidth]{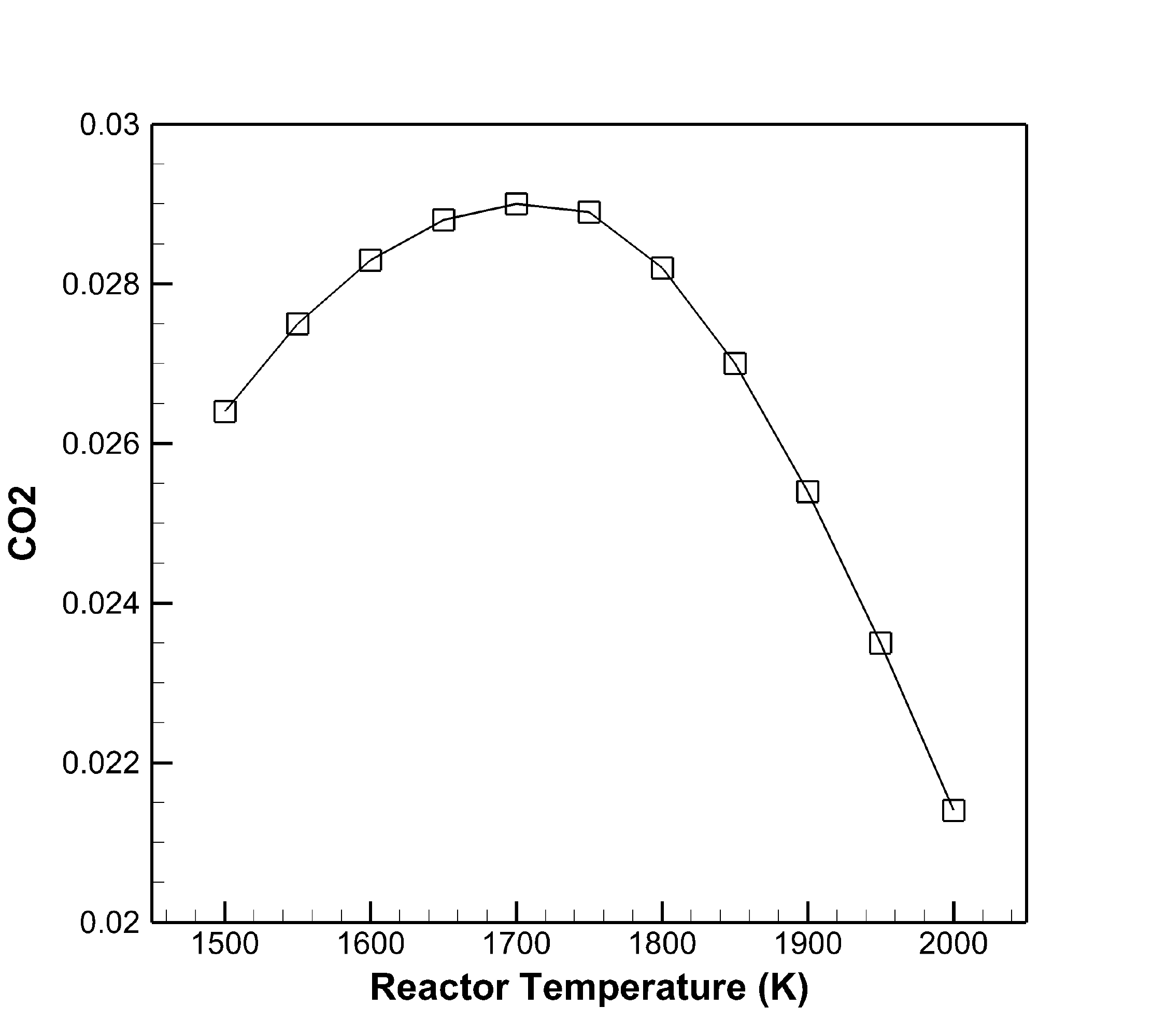}}
  \subfigure{
  \includegraphics[width=0.45\textwidth]{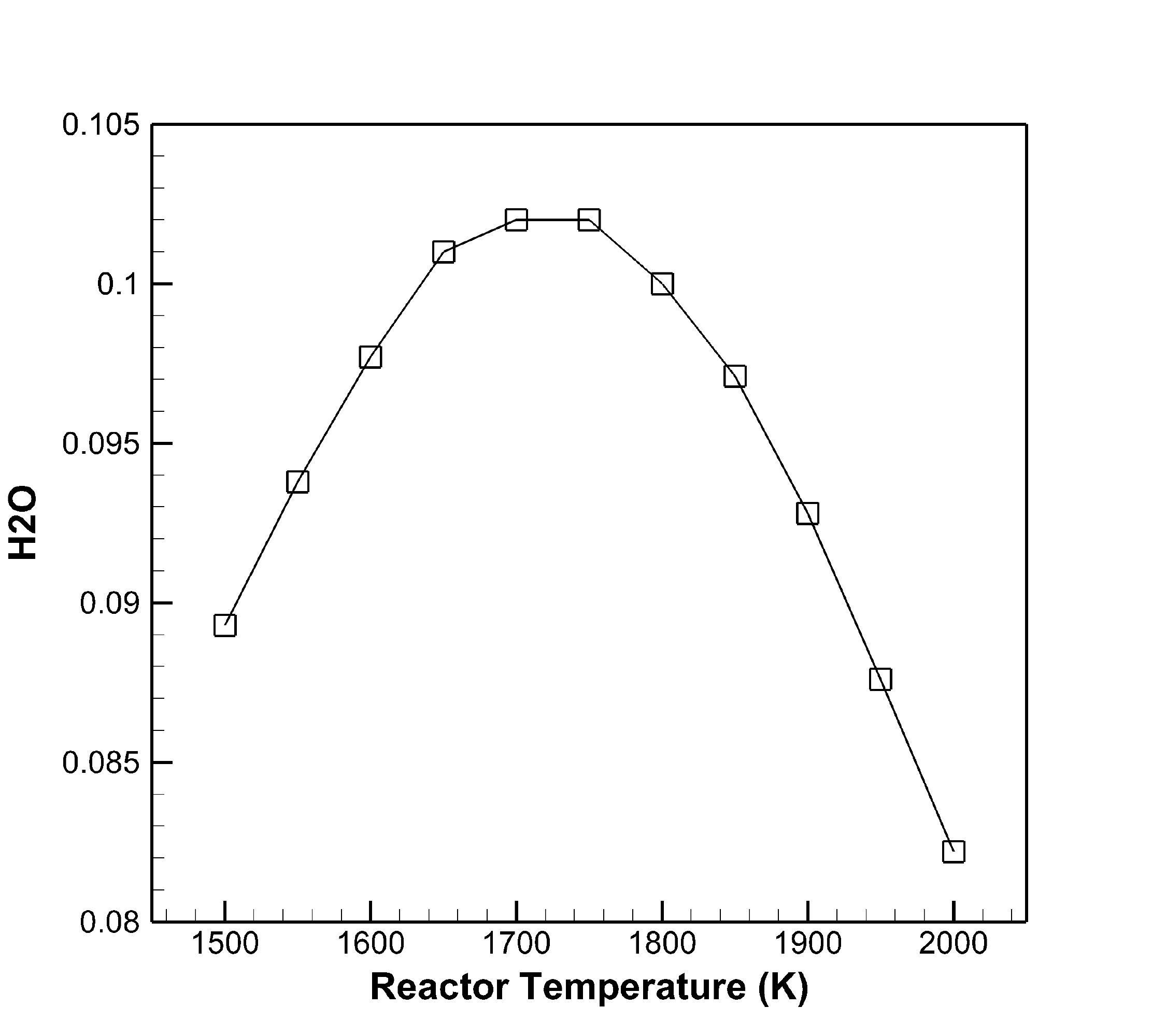}}  
  \caption{Mole fraction of C2H4 (top left), Mole fraction of O2 (top right), Mole fraction of CO2 (bottom left), Mole fraction of H2O (bottom right) with reactor temperature for ethylene/air combustion from PSR simulations; GPU Result with MOMIC - lines with open square symbols }
  \label{fig:Species_mole_fraction_vs_TEMP}
\end{figure}

The next two figures show the behavior of major and minor gaseous species with respect to residence time and reactor temperature.  Figure \ref{fig:Species_mole_fraction_vs_TAU} shows the variation of the mole fraction of $\textrm{C}_{2}\textrm{H}_{4}$, $\textrm{O}_{2}$, $\textrm{C}\textrm{O}_{2}$, and $\textrm{C}\textrm{H}_{4}$ with reactor residence time.  $\textrm{C}_{2}\textrm{H}_{4}$ and $\textrm{O}_{2}$ decrease with increasing residence time, which is expected  because of the fact that higher the residence time higher will be the time available for the reactants to get consumed and be converted into products.  The $\textrm{C}\textrm{O}_{2}$ concentration as a function of residence time exhibited a steep increase followed by almost constant value while $\textrm{C}\textrm{H}_{4}$ increase initially to a maximum at around 30 ms and then decreases continuously.  Figure \ref{fig:Species_mole_fraction_vs_TEMP} shows the variation of mole fraction of gaseous species with respect to reactor temperature. $\textrm{C}_{2}\textrm{H}_{4}$ is being consumed at higher rate for the temperature range of 1500 K to 1850 K as compared to the temperature range of 1850 K to 2000 K. As anticipated, consumption of $\textrm{O}_{2}$ is increasing with respect to reactor temperature. Concentration of 
$\textrm{C}\textrm{O}_{2}$ and $\textrm{H}_{2}\textrm{O}$ exhibited an increase to a maximum at 1750 K and then a continuous decrease.

\section{Speedup}

The GPU implementation of a batched PSR calculation presented here is integrated into our in-house PSR solver, which in turn is based on a hybrid/newton time integration approach \cite{adhikari2016hybrid}.  A batched PSR calculation (of say $n$ PSRs) refers to a set of $n$ PSRs that are solved simultaneously.  The performance of the GPU implementation over the CPU is assessed by determining the speedup, which is the ratio of the CPU computation time over the GPU computation time.  CPU computations were performed on a workstation with Intel(R) Xeon(R) E5-2643 3.30 GHz processor while GPU computations were performed on NVIDIA Tesla K40. The speedup being shown here is for PSR batches for a single newton iteration only. 

\begin{figure}[h]
  \centering
  \includegraphics[width=1\textwidth]{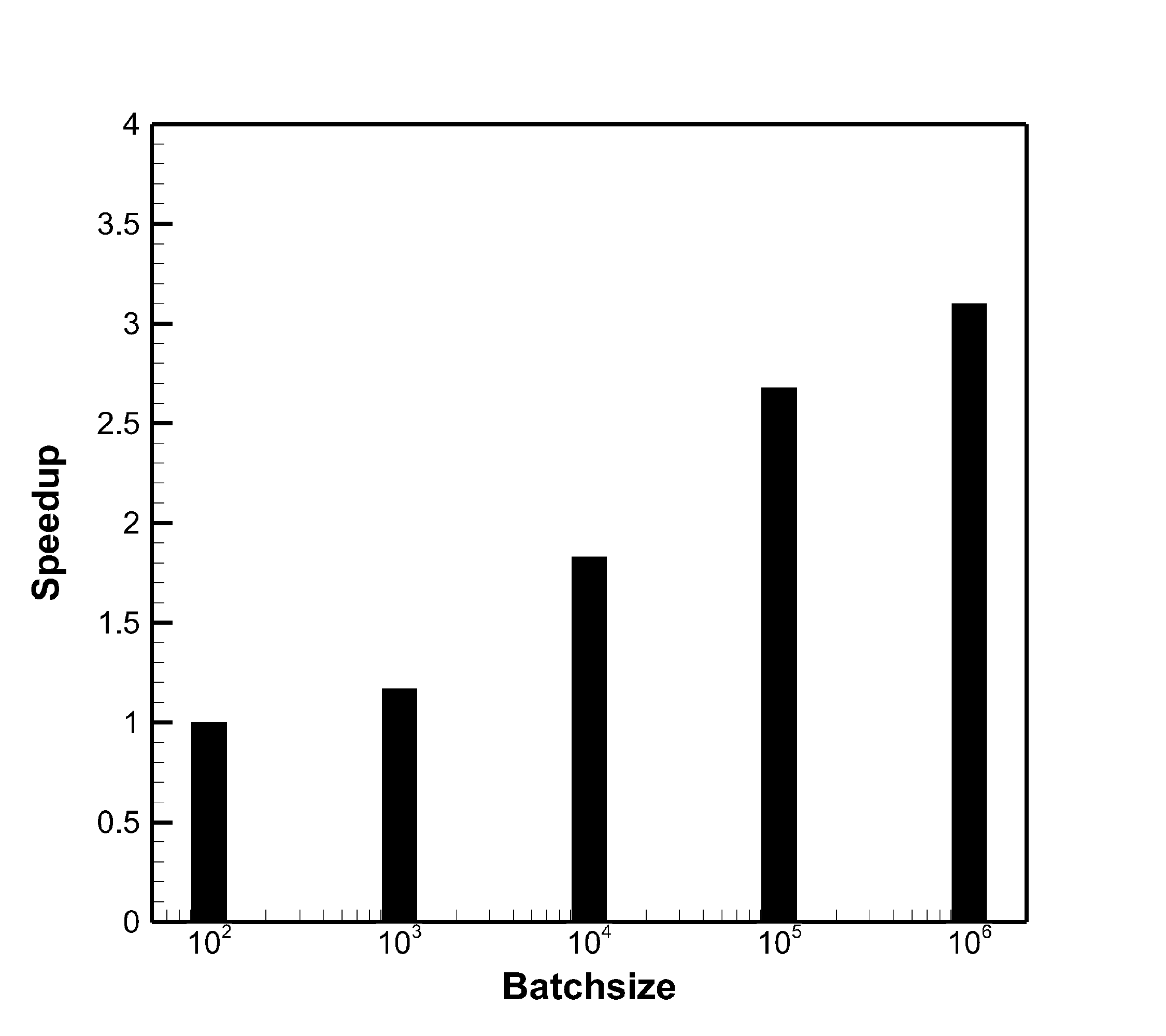}
  \caption{Speedup versus size of batched PSR  }
  \label{fig:speedup_batched_psr}
\end{figure}

CPU and GPU calculations for ethylene combustion in PSR are carried out for a range of PSR batch sizes from 100 to 1 million, with each of the PSR having a constant temperature of 1800K and residence time of 50 ms. Inlet composition of each PSR in the batch is defined by the mole fraction of reacting species; 0.15 mole of $\textrm{C}_{2}\textrm{H}_{4}$, 0.18 mole of $\textrm{O}_{2}$, and 0.67 mole of $\textrm{N}_{2}$. The reaction mechanism used in the PSR calculations consists of 101 species and 542 reactions \cite{JHM2000}. 
Figure \ref{fig:speedup_batched_psr} shows the GPU speedup for the different batch sizes.  While the GU and CPU timings are comparable for batch sizes below $10^4$, the maximum speedup obtained was 3 for a batch size of 1 million.
%


\section{Conclusion}
A highly-parallelized GPU algorithm for solving a batched PSR was developed, coupled to the moment-based soot modeling method, MOMIC, and successfully tested for ethylene-combustion. The GPU algorithm is based on the GPU-enhanced algorithm for forward reaction rate calculation, equilibrium constant calculation, and reverse rate constant evaluation.  This algorithm was first validated by comparing GPU and CPU results from the current PSR development to previous PSR calculations of ethylene combustion \cite{NKM1998} for various reactor temperatures and residence times.  Excellent agreement was observed in these comparisons, indicating the correctness of the implementation.  In addition, the GPU-enhanced algorithm was tested for 5 different batchsizes and GPU algorithm was found to be three times faster than the CPU algorithm for PSR batchsize of one million. The algorithm that has been presented here can be applied to other combustion simulations that make use of detailed kinetics.

\bibliographystyle{wssci}
\bibliography{PSR_GPU_BATCHED}

\end{document}